\documentclass[aps,prd,twocolumn]{revtex4}

\usepackage{aas_macros}
\usepackage{natbib}
\usepackage[latin1]{inputenc}
\usepackage{amsfonts}
\usepackage{amsmath}
\usepackage{amssymb}
\usepackage[T1]{fontenc}
\usepackage[dvips]{graphicx}
\newcommand{\hs}{\hspace{15pt}}
\newcommand{\dv}{\textrm{d}}

\newcommand{\be}{\begin{equation}}
\newcommand{\ee}{\end{equation}}
\newcommand{\ba}{\begin{eqnarray}}
\newcommand{\ea}{\end{eqnarray}}
\newcommand{\ch}{\mathcal{H}}
\newcommand{\fot}{\frac{1}{2}}
\newcommand{\gmn}{g_{\mu\nu}}

\newcommand{\partis}{\partial_i\partial^i}

\newcommand{\sg}{\sqrt{-g}}

\newcommand{\kap}{\frac{1}{16\pi G}}
\newcommand{\npsis}{(\nabla\psi)^2}

\newcommand{\tio}{\tilde\Omega}

\begin{document}

\title{The Distinguishability of Interacting Dark Energy from Modified Gravity}

\author{Timothy Clemson, Kazuya Koyama}

\affiliation{Institute of Cosmology \& Gravitation, University of Portsmouth, Portsmouth PO1 3FX, United Kingdom}

\date{\today}

\begin{abstract}
We study the observational viability of coupled quintessence models with their expansion and growth histories matched to modified gravity cosmologies. We find that for a Dvali-Gabadadze-Porrati model
which has been fitted to observations, the matched interacting dark energy models are observationally disfavoured. We also study the distinguishability of interacting dark energy models matched to
scalar-tensor theory cosmologies and show that it is not always possible to find a physical interacting dark energy model which shares their expansion and growth histories.
\end{abstract}

\maketitle

\section{Introduction}
The Universe appears to be undergoing a late-time accelerated expansion~\cite{rie98,per99}. A model which includes a cosmological constant $\Lambda$ and cold dark matter~(CDM) evolving according to
Einstein's theory of General Relativity~(GR) provides the best description of this~\cite{kom11}. There are many alternative explanations however, the two main classes of which are modified
gravity~(MG), (see~\cite{cap11,noj11,cli12} for reviews), and dark energy~(DE), (see~\cite{cop06,fri08,li11} for reviews), and we must rely on observations to discriminate between them~\cite{alb06}.

It is always possible to find a DE model with a time varying equation of state parameter which produces a given expansion history~\cite{cap05,noj06,son07}, so in a worst-case
scenario a DE model could exactly mimick a MG model's expansion history, making them indistinguishable. To break this degeneracy it is necessary to take differences in the growth of structure into
account and a great deal of effort has gone into distinguishing DE from
MG~\cite{lin05,ish06,kno06,nes06,pol06,chi07,hea07,hut07,lin07,uza07,wan07,yam07,acq08,ame08,las08,wan08,hu09,wu09,bag10,che10,hut10,jin10,sha10,sim10,sim11,lee11,jen11,wan11,zia12}. It has been
argued that by finetuning the properties of a DE model its structure growth can also be made to mimick that of a given MG theory~\cite{kun07,ber08,sap10}, but by employing suitable
combinations of observables consistency tests can be made which should be able to distinguish between realistic models~\cite{jai08,son09}.

The above works focus on minimally coupled DE but it's also possible to match the growth and expansion histories of MG with interacting dark energy~(IDE)
models~\cite{wei08}, (for recent IDE works see~\cite{cle12} and references therein). IDE models can look like modifications of GR~\cite{hon10}, but they should deviate from GR+$\Lambda$CDM in a
way which is distinct to that of MG~\cite{son10}. In this paper we investigate their distinguishability by testing the observational viability of IDE models with their growth and
expansion histories matched to MG cosmologies, restricting ourselves to a flat spacetime in the Newtonian regime.

Section~\ref{DGP} revisits an example Dvali-Gabadadze-Porrati~(DGP)~\cite{dva00} model from~\cite{wei08} to examine the observational distinguishability of matched IDE/DGP models. Section~\ref{STT}
extends the matching procedure used for the DGP case to a more general scalar-tensor theory~(STT) model and again considers whether the matched IDE models can be distinguished from their MG
counterparts observationally. Our conclusions are then drawn in section~\ref{conclusions}.
\section{Interacting dark energy matched to a DGP cosmology}\label{DGP}
For a scalar field model of IDE a general action may be written as,
\ba S_{IDE}&=&\kap\int d^4x\sg\left[R-\fot\npsis-V(\psi)\right]\nonumber\\&&+S_m(\gmn,\psi,\varphi),\label{ideaction}\ea
where $S_m(\gmn,\psi,\varphi)$ is the matter action, with $\varphi$ being the matter field. In~\cite{wei08} the authors matched a generalised IDE model to a particular choice of DGP model which had
been fitted to observations. They used the IDE potential and coupling functions to match the DGP expansion and growth histories respectively. Essentially the evolution of the background CDM
density $\rho$ in the IDE model is determined by the matching of its perturbation $\delta\equiv\delta\rho/\rho$ to that of the DGP model.

\begin{figure*}
 \includegraphics[width=\columnwidth]{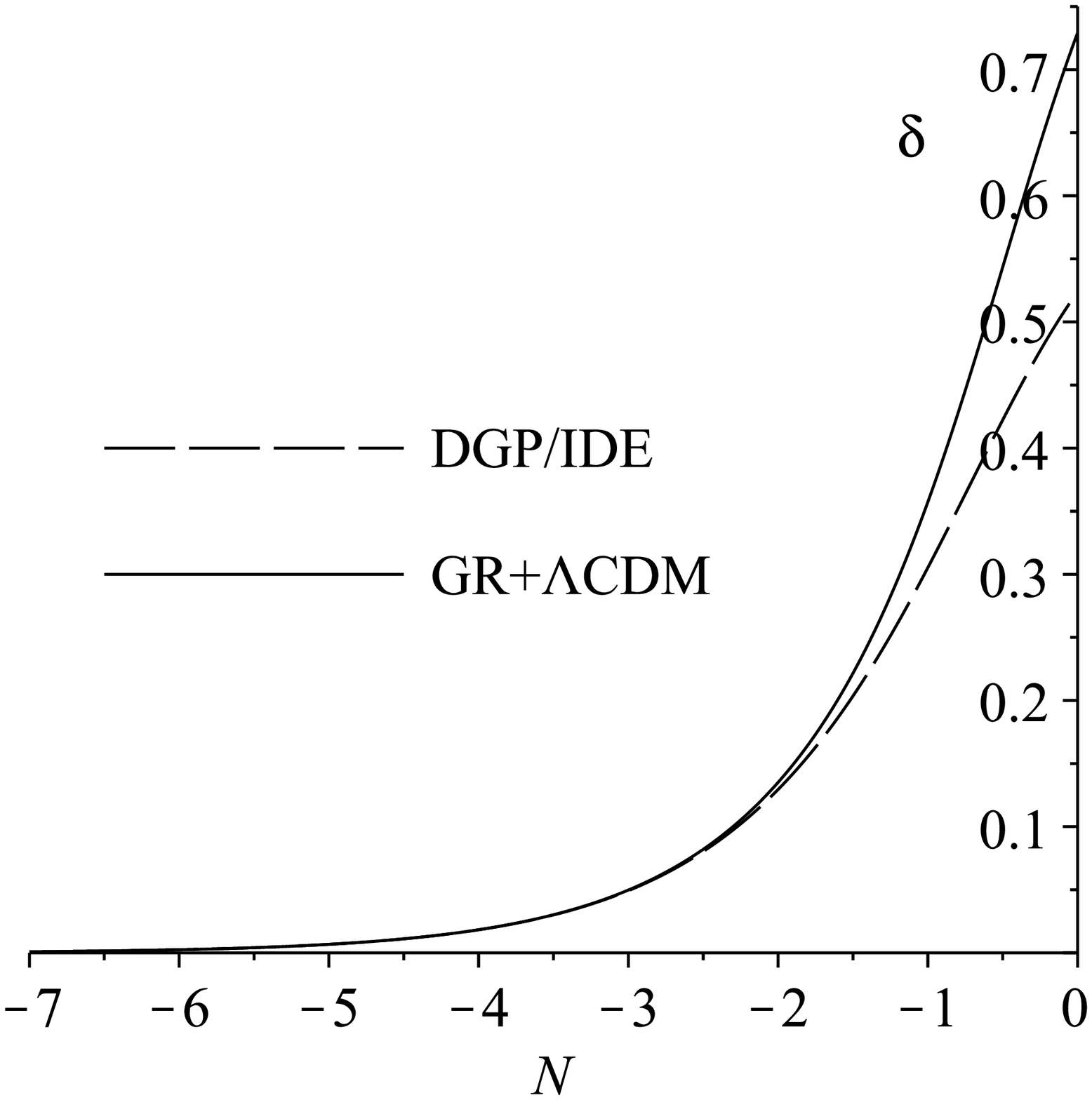}
 \includegraphics[width=\columnwidth]{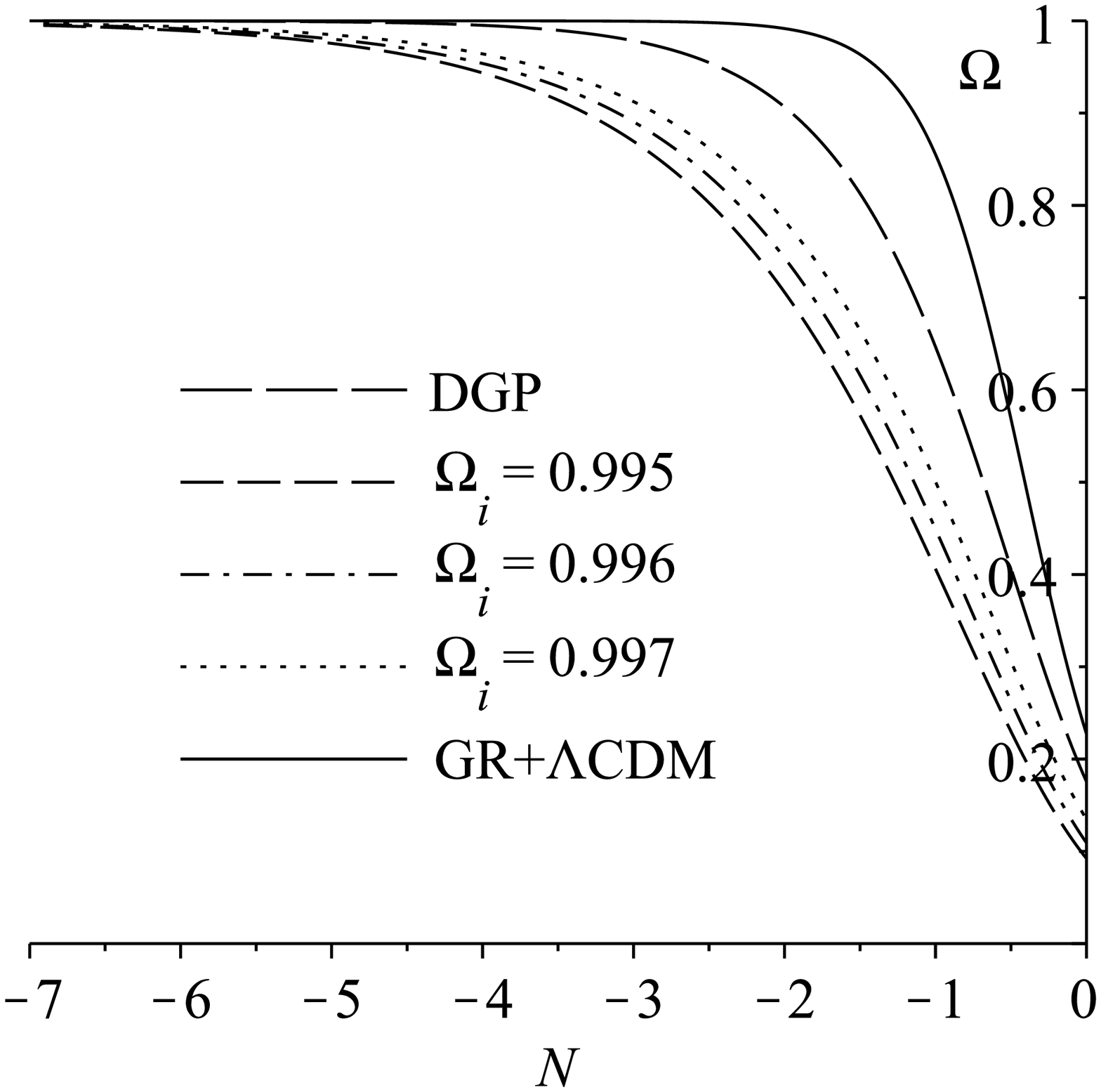}
 \caption{Evolution of the density perturbation (left) and the density parameters (right) for the matched DGP/IDE models, each with a different $\Omega_i$, and a GR+$\Lambda$CDM model.
\label{matched}}
 \end{figure*}

Fig.~\ref{matched} shows the evolution of $\delta$
and the different background evolutions of the IDE and DGP density parameters $\Omega$ and $\tilde\Omega$, (tildes denote MG quantities throughout), where
$\Omega\equiv8\pi Ga^2\rho/(3\ch^2)$ with $\ch\equiv a^{-1}\dv a/\dv\tau$ and $\tau$ is conformal time. Also plotted for comparison is a GR+$\Lambda$CDM model chosen to give $\Omega_0\approx0.227$,
(subscript $0$'s denote present day quantities throughout), in line with recent constraints~\cite{kom11}.

The original example had initial conditions set early in the matter dominated era with an initial DGP energy density parameter
$\tilde\Omega_i\approx1$, (subscript $i$'s denote initial values). The initial IDE density parameter was $\Omega_i=0.995$ and in
addition to this solution we plot the result of choosing $\Omega_i=0.996$ and $\Omega_i=0.997$ in Fig.~\ref{matched}, but find that there are no solutions with
$\Omega_i\gtrsim=0.997$, (see Appendix~\ref{dgpiclimit}).

This means that there is a limit on how closely one can hope to match the evolution of the IDE/DGP densities through the choice of the
boundary conditions on $\Omega$. This difference should be evident in any quantity which depends on the CDM density, for example the sum of the metric potentials.

\begin{figure*}
 \includegraphics[width=\columnwidth]{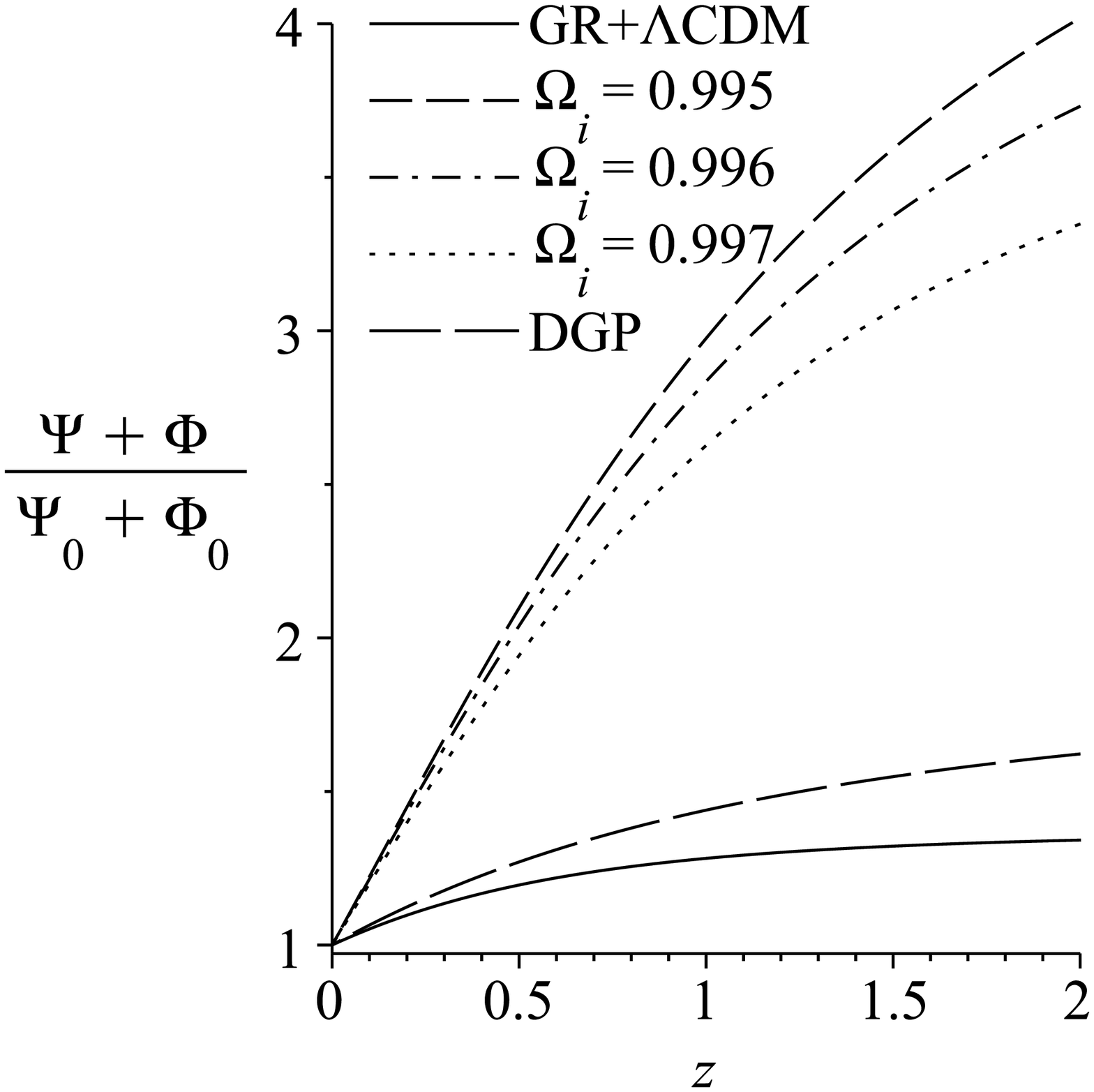}
 \includegraphics[width=\columnwidth]{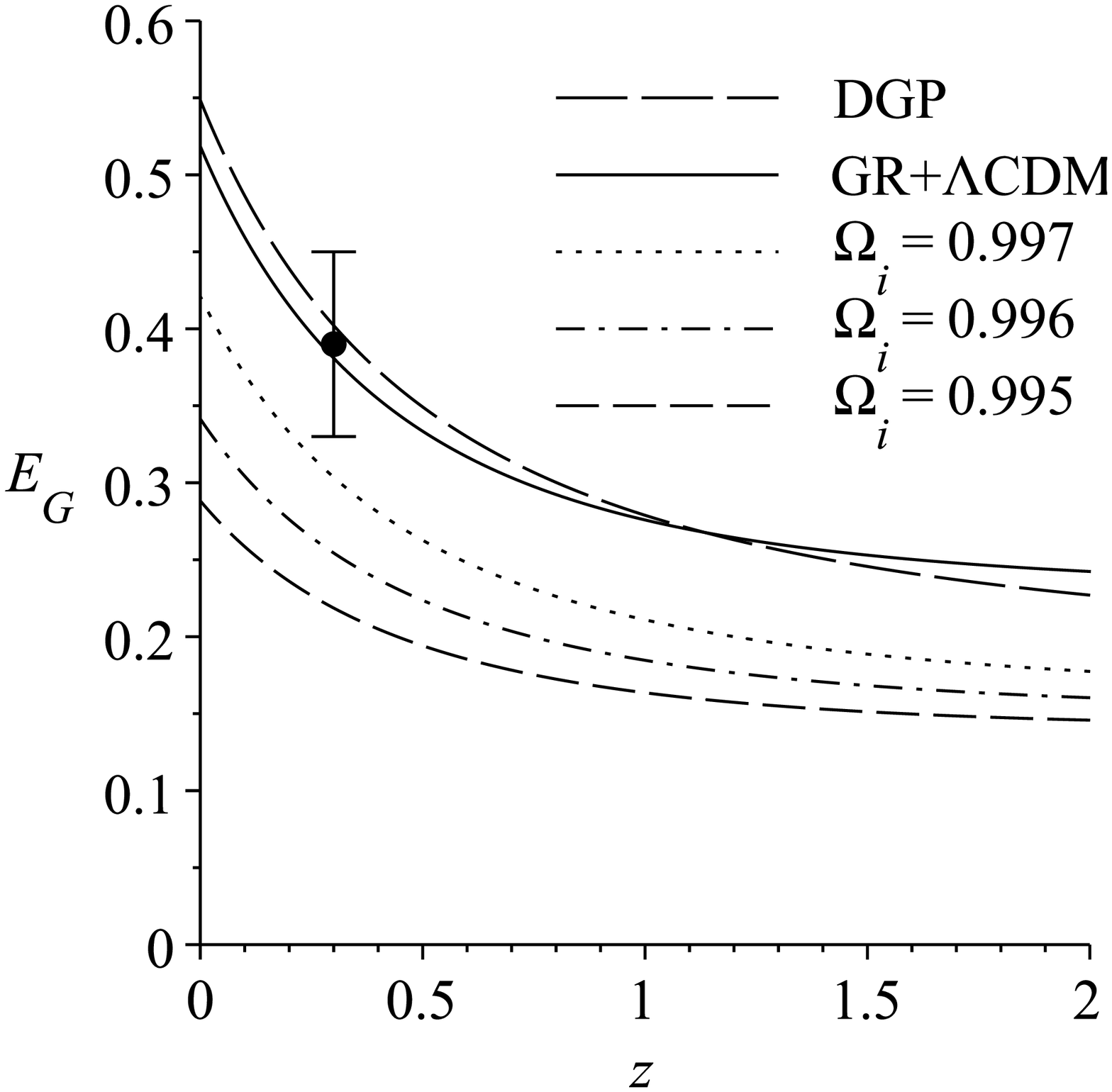}
 \caption{Evolution of the sum of the metric potentials normalised at the present day (left) and the $E_G$ parameter (right) for the matched DGP/IDE models, each with a different $\Omega_i$, and a
GR+$\Lambda$CDM model. The observational measurement is $E_G=0.39\pm0.06(1\sigma)$ at an effective redshift of $z=0.3$~\cite{rey10}.\label{dgppotentials}}
 \end{figure*}

The perturbed metric in the Newtonian regime may be written,
\be ds^2=a^2[-(1+2\Psi)d\tau^2+(1-2\Phi)\gamma_{ij}dx^idx^j].\ee
Both the DGP and IDE models obey the same evolution equation for the sum of the metric potentials $\Psi$ and $\Phi$,
\be \partial_i\partial^i(\Psi+\Phi)=8\pi Ga^2\tilde\rho\delta,\hs \partial_i\partial^i(\Psi+\Phi)=8\pi Ga^2\rho\delta.\label{dgpsum}\ee
This quantity is plotted in the left-hand panel of Fig.~\ref{dgppotentials} as a function of redshift at late times, making clear the significant distinction arising between the IDE and DGP models
from the restriction on the boundary conditions for $\Omega$.

One way to test for this difference observationally is to use the $E_G$ prameter~\cite{zha07} defined by,
\be E_G\equiv\left[\frac{\partial_i\partial^i(\Psi+\Phi)}{-3\ch_0^2a^{-1}\theta}\right]_z,\label{eg}\ee
where in the Newtonian regime $\theta=-\delta'$, (primes denote derivatives with respect to $N\equiv\ln(a)$ throughout). Note however that this relation does not hold for
all IDE models, eg.~\cite{cle12}. The numerator in Eq.~(\ref{eg}) can be measured from weak lensing observations, while the denominator can be found from peculiar velocity measurements and for the
models studied here we have,
\ba E_G^{DGP}=\frac{\tilde\Omega_0\delta}{\delta'},\label{dgpeg}
\\ E_G^{IDE}=\frac{a\ch^2\Omega\delta}{\ch^2_0\delta'}\label{ideeg}.\ea
The right-hand panel of Fig.~\ref{dgppotentials} shows $E_G$ at late times for the DGP, IDE and GR+$\Lambda$CDM models along with some recent observational constraints
\cite{rey10}. We can see that although the DGP model is a good fit, even the worst-case IDE model with boundary conditions as close as possible to those of the DGP
model is disfavoured by observations.
\section{Interacting dark energy matched to scalar-tensor theory cosmologies}\label{STT}
In the same vein as the previous section, we now apply a similar method to a simple STT model in order to explore the potential distinguishability for particular parameter
values. We assume it to be the large scale limit of MG models to which local constraints on the gravity theory~\cite{wil06} do not apply due to a screening mechanism such as the
chameleon~\cite{kho04,nav07,fau07}, thus allowing the effect of baryons to be neglected.

The action for a STT model may be written as,
\ba S_{STT}&=&\kap\int d^4x\sg\left[\phi  R-\frac{\omega}\phi(\nabla\phi)^2-U\right]\nonumber\\&&+S_m(\gmn,\varphi).\ea
In general $\omega$ and $U$ can be functions of $\phi$ but for our purposes we take them to be constant. The
acceleration, scalar field and density perturbation equations derived from this action are,
\ba \ch'=-\frac{\ch}2\left(\frac{\tio}{\phi}+\frac{\ch'}{\ch}\frac{\phi'}{\phi}+\frac{\phi''}{\phi}\right)-\frac{\ch\omega}3\left(\frac{\phi'}\phi\right)^2+\frac{Ue^{2N}}{
6\ch\phi}\label{mgacc},
\\ \phi''=-\left(2+\frac{\ch'}\ch\right)\phi'+\frac1{2\omega+3}\left(3\tio+\frac{2Ue^{2N}}{\ch^2}\right),
\\\delta''=-\left(1+\frac{\ch'}\ch\right)\delta'+\frac{3\tio\delta}{2\phi}\left(1+\frac1{2\omega+3}\right).\label{mgdelta}\ea
These equations determine the expansion and growth histories for both the STT and
matched IDE models. The action for the IDE model, Eq.~(\ref{ideaction}), leads to its fluid, scalar field, Friedmann and density perturbation equations,
\ba\rho'=-3\rho-\fot C\rho\psi',\label{idefluid}\\
\psi''=-\left(2+\frac{\ch'}{\ch}\right)\psi'-\frac{a^2V'}{\ch^2\psi'}+\frac{Ca^2\rho}{2\ch^2},\\
\ch^2=\frac{8\pi G}3a^2\left(\rho+\frac{\ch^2\psi'}{2e^{2N}}+V\right),\label{idefriedmann}\\
\delta''=-\left(1+\frac{\ch'}{\ch}-\fot C\psi'\right)\delta'+\frac32\Omega\delta\left(1+\frac{C}{16\pi G}\right),\label{idedelta}\ea
where $C$ is the DE/CDM coupling function and $V$ is the scalar field potential, both of which are taken to be free functions. Using Eq's~(\ref{idefluid}-\ref{idefriedmann}) and comparing
Eq.~(\ref{mgdelta}) to Eq.~(\ref{idedelta}) now leads to a differential equation for $\Omega$,
\ba\fot \left(1+\frac{\Omega'}\Omega+2\frac{\ch'}\ch\right)\delta_c'&=&-\frac32 \delta_c\bigg[\left(1+\frac{1}{2\omega+3}\right)\tio\nonumber\\&&-\left(1+\frac{C^2}{16\pi
G}\right)\Omega\bigg],\label{matchedeq}\ea
where,
\be C^2=16\pi G\frac{\left(1+\frac{\Omega'}\Omega+2\frac{\ch'}\ch\right)^2}{1-\frac{\ch'}\ch-\frac32\Omega}.\label{c2}\ee
Eq.~(\ref{matchedeq}) is quadratic in $\Omega'$ and so we choose the root which is typically negative initially, (the alternative branch typically leads to increasing $\Omega$ and the limits
described in the Appendices are reached before the present day).  We can now solve Eq's~(\ref{mgacc}-\ref{mgdelta}) numerically, along with the root of Eq.~(\ref{matchedeq}),
to find $\Omega$, $\ch$, $\delta$, $\delta'$, $\phi$ and $\phi'$ at any given $N$.
In this way the freedom in the coupling function $C$ is explicitly used to match the evolutions of the $\delta$'s, while the freedom in the scalar field potential $V$ is used implicitly to match
the expansion histories via the IDE Friedmann constraint, Eq.~(\ref{idefriedmann}). The initial conditions used are,
\ba N_i=-7,\hs \phi_i=1,\hs \phi_i'=0,\nonumber\\ \ch_i=1, \hs \delta_i=a_i,\hs \delta_i'=a_i.\ea
\begin{figure*}
 \includegraphics[width=\columnwidth]{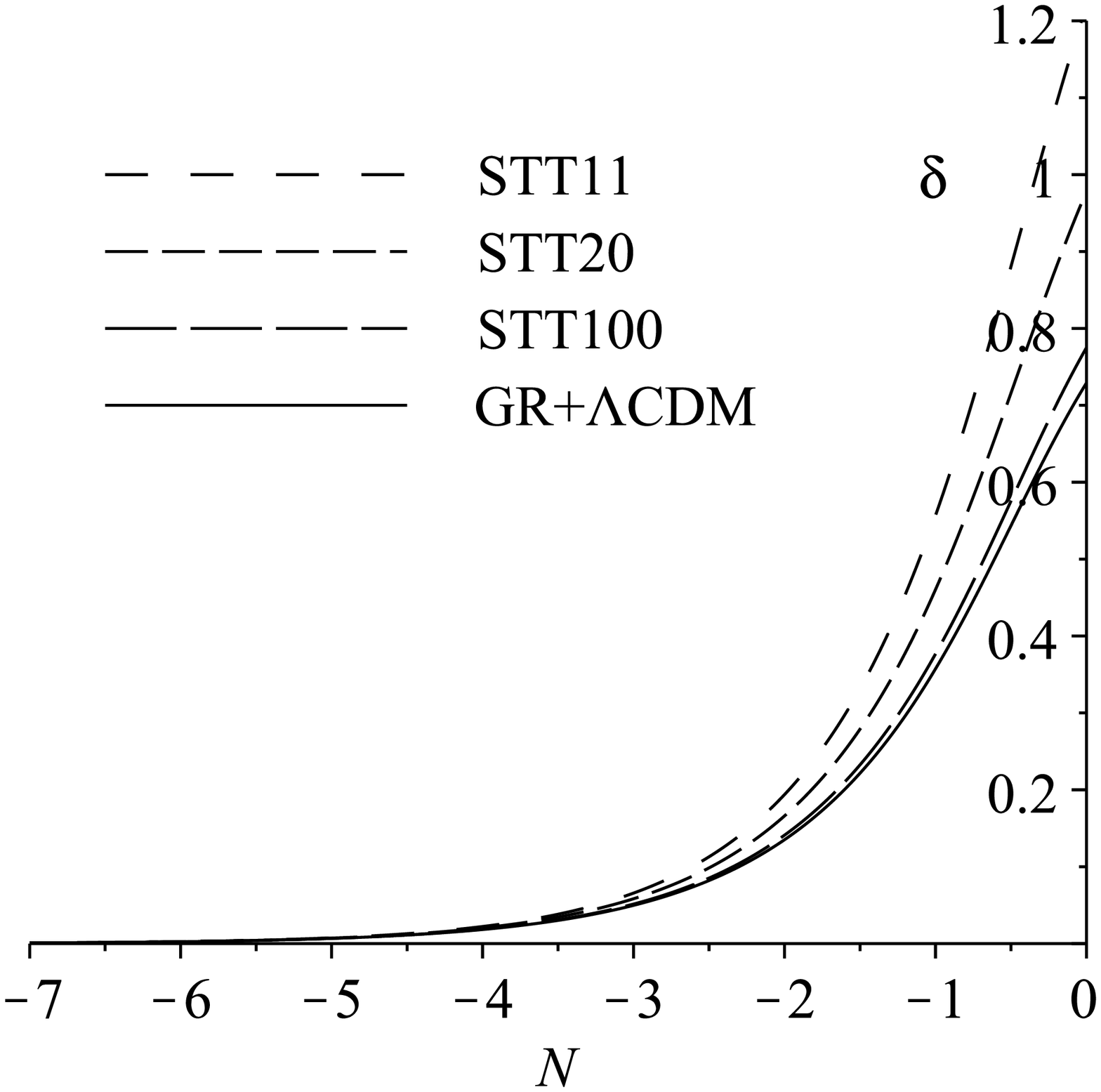}
 \includegraphics[width=\columnwidth]{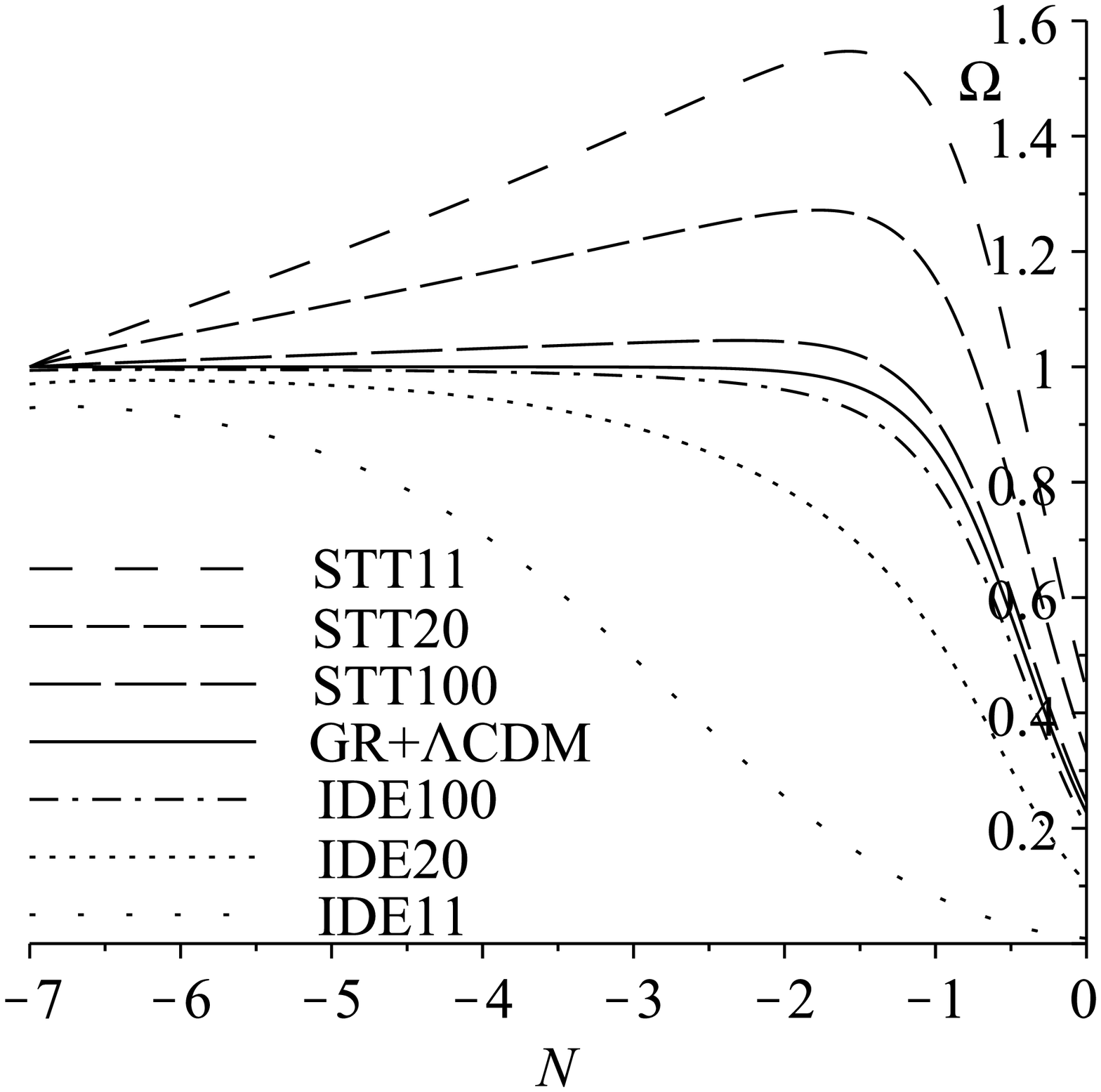}
 \caption{Evolution of the density perturbation (left) and the density parameters (right) for the matched STT/IDE models and a GR+$\Lambda$CDM model. The models STT11, STT20 and STT100 have
$\omega=11$, $\omega=20$ and $\omega=100$ respectively, with IDE11, IDE20 and IDE100 being their matched IDE counterparts. Note that including $\phi$ in the definition of $\tilde\Omega$ would bring
the STT models on the right much closer to the GR+$\Lambda$CDM $\Omega$ curve. Even then however $\tilde\Omega>1$ is still possible despite the spacetime being flat because the sum of the
gravitational scalar field terms in the STT `Friedmann' equation can be negative.\label{stmatched}}
\end{figure*}
$\tilde\Omega_i$ is chosen so that $\tilde\Omega_0$ is the same as the previously mentioned GR+$\Lambda$CDM model's present day density parameter when $\omega\to\infty$, while $U_i$ is
determined by the choice of $\tilde\Omega_i$ due to the `Friedmann' constraint,
\be U_i=\frac{6\ch_i^2}{a_i^2}\left(\phi_i-\tilde\Omega_i+\phi_i'-\frac{\omega\phi_i'^2}{6\phi_i}\right).\ee
As in the case of the earlier DGP example there is a limit on how close $\Omega_i$ can be to $\tilde\Omega_i$ (see Appendix~\ref{stlimitappendix}). Fig.~\ref{stmatched} shows results for three
different values of $\omega$ where in each case $\Omega_i$ has been chosen to be as close as possible to $\tio_i$ in the spirit of representing a worst-case scenario for distinguishing between the
IDE/STT models.
\begin{figure*}
 \includegraphics[width=\columnwidth]{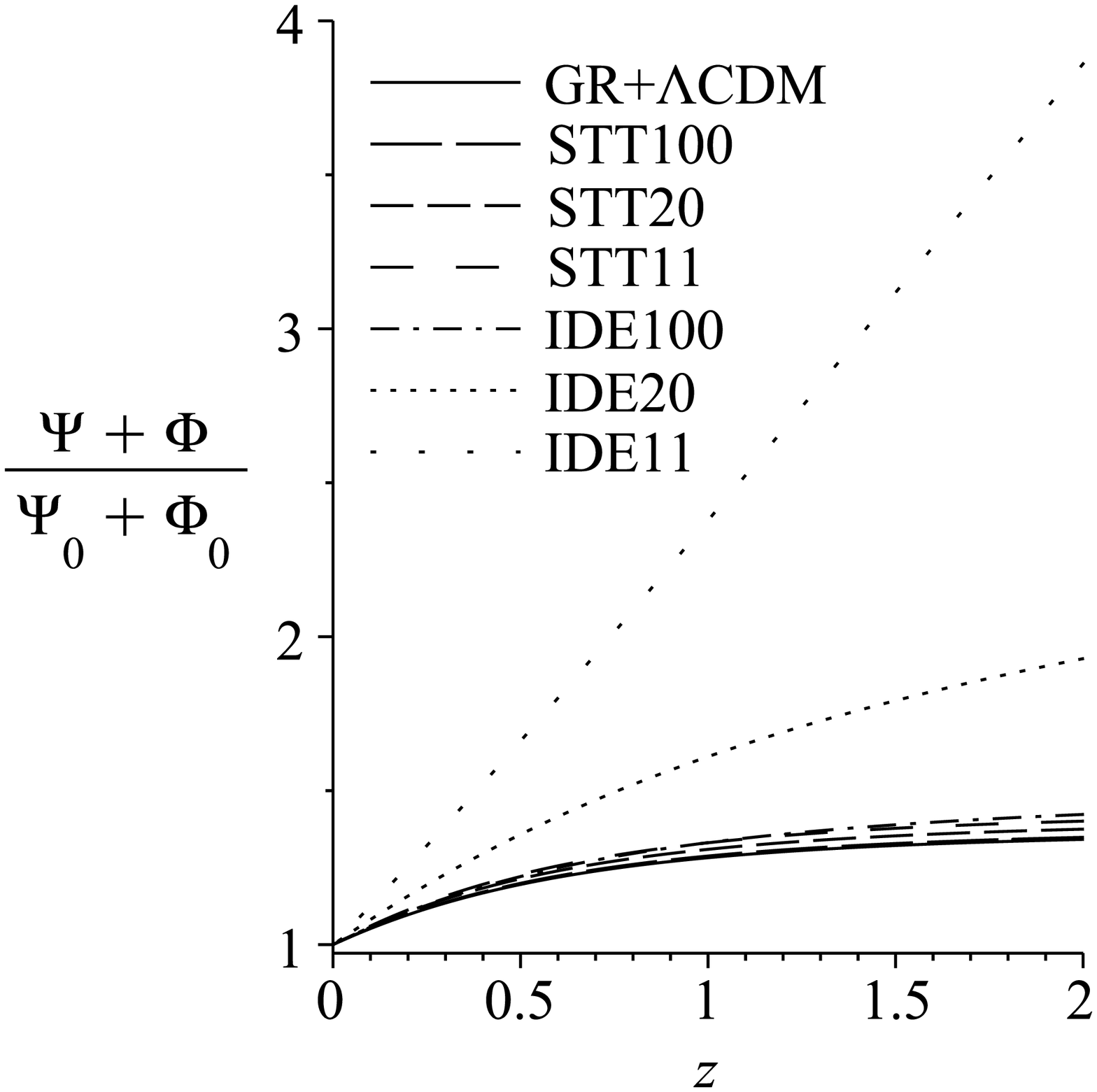}
 \includegraphics[width=\columnwidth]{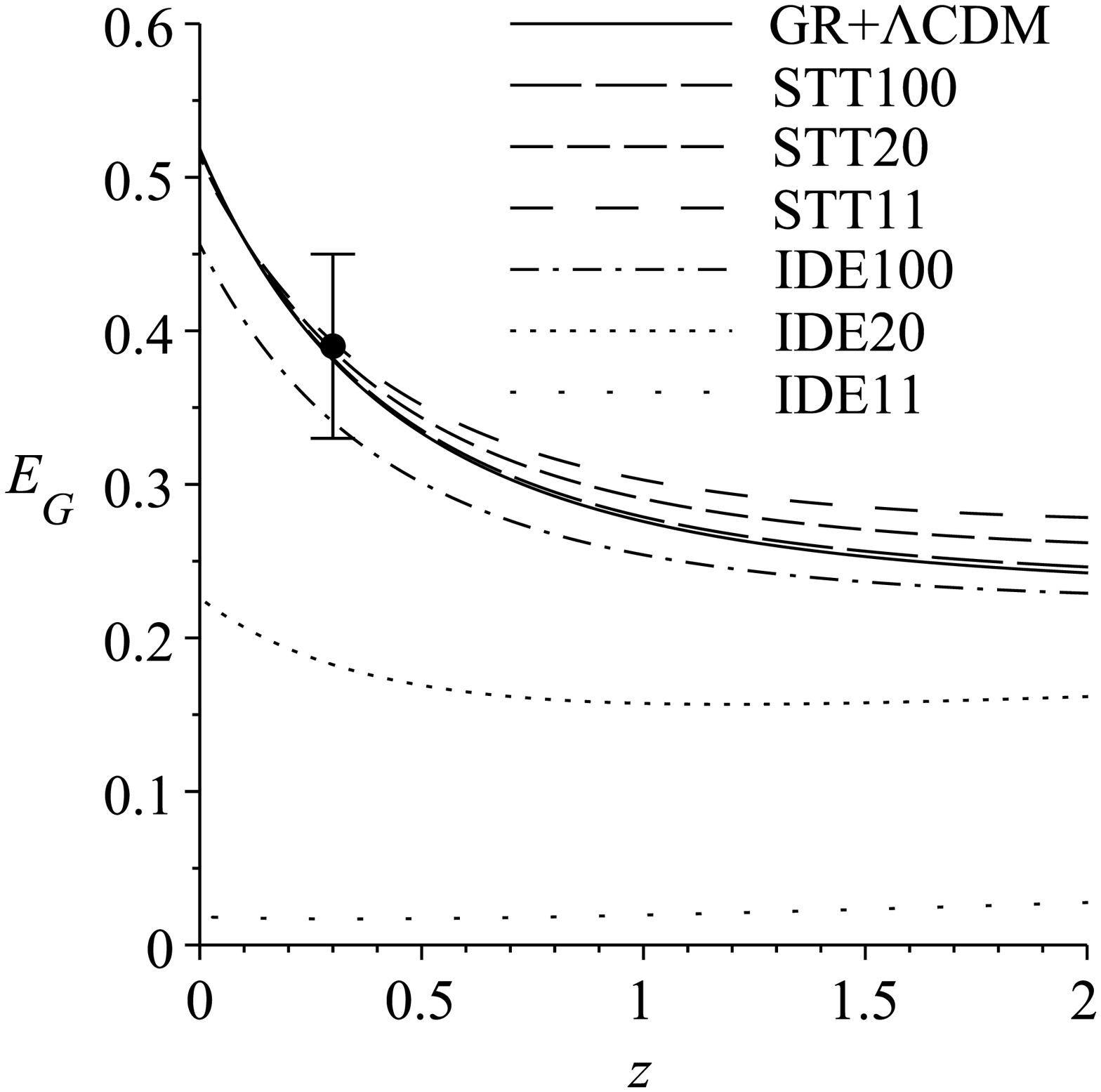}
 \caption{Evolution of the sum of the metric potentials normalised at the present day (left) and the $E_G$ parameter (right) for the matched STT/IDE models, each with a different $\Omega_i$, and a
GR+$\Lambda$CDM model. The models STT11, STT20 and STT100 have $\omega=11$,
$\omega=20$ and $\omega=100$ respectively, with IDE11, IDE20 and IDE100 being their matched IDE counterparts. The observational measurement is $E_G=0.39\pm0.06(1\sigma)$ at an effective redshift of
$z=0.3$~\cite{rey10}. \label{stpotentials}}
 \end{figure*}
The evolution equation for the sum of the metric potentials in the STT model is,
\be  \partis(\Psi+\Phi)=\frac{8\pi \tilde G}{\phi}a^2\tilde\rho\delta,\label{stsum}\ee
where $\tilde G=G\phi_0$, leading to,
\be E_G^{STT}=\frac{\tilde\Omega_0\delta}{\phi\delta'},\label{steg}\ee
with the IDE expression as before in Eq.~(\ref{ideeg}). Fig.~\ref{stpotentials} shows $\Psi+\Phi$ and $E_G$ as functions of $z$ at late times for the STT models and their matched IDE counterparts.
Once again the IDE models lie much farther from the GR+$\Lambda$CDM case than their MG counterparts, with all but that matched to the $\omega=100$ STT model lying outside of the observational
constraints on $E_G$.

In~\cite{tsu08} it was shown that constraints on STT models from cosmic microwave background, matter power spectrum and local gravity measurements could be avoided
using a chameleon mechanism, leading to only a weak bound of $\omega>-1.28$. Our model here is essentialy a Brans-Dicke theory~\cite{bra61} plus a cosmological constant, for which lower bounds of
$\omega>120(2\sigma)$~\cite{acq05} and $\omega>97.8(2\sigma)$~\cite{wu10} have been found, (Note that~\cite{nag04} give a lower bound of $\omega>1000(2\sigma)$, but see discussions
in~\cite{acq05,wu10}).

The addition of supernova data would significantly improve constraints on $\omega$, but account would need to be taken of local~\cite{cli05}
and temporal~\cite{gaz02} variation in the gravitational scalar field $\phi$. In~\cite{acq07} a recovery of GR at late-times sufficient to allow the use of supernova data was assumed and bounds of
$\omega>500-1000$ from future data were forecast. If these constraints can be acheived it will not be possible to distinguish between our matched STT/IDE models with the $E_G$ results we use here,
although with new data of course the $E_G$ constraints could also be tightened.
\section{Conclusions}\label{conclusions}
We have shown that although it is possible to construct an IDE model which matches the growth and expansion histories of a DGP model fitted to observations, even in the worst-case scenario, where
their density evolutions are as close as theoretically possible, the matched IDE model can be distinguished by observations. 

For our simple STT model and its matched IDE counterpart we have calculated a limit on how similar the initial matter densities can be. This limit depends on
the strength of deviation from GR and we find that in cases which differ significantly from GR+$\Lambda$CDM even the worst-case matched IDE model can be distinguished by observations.

We have also shown that it is not always possible to construct a physical IDE model which matches the growth and expansion histories of our STT models and that there is a
limit on the strength of deviation from GR, beyond which the time derivative of the IDE scalar field becomes complex before the present day.
\section{Acknowledgements}
TC was funded by a UK Science \& Technology Facilities Council (STFC) PhD studentship. KK is supported by the STFC (grant no. ST/H002774/1), the European Research Council and the Leverhulme
trust.
\appendix
\begin{figure*}
 \includegraphics[width=\columnwidth]{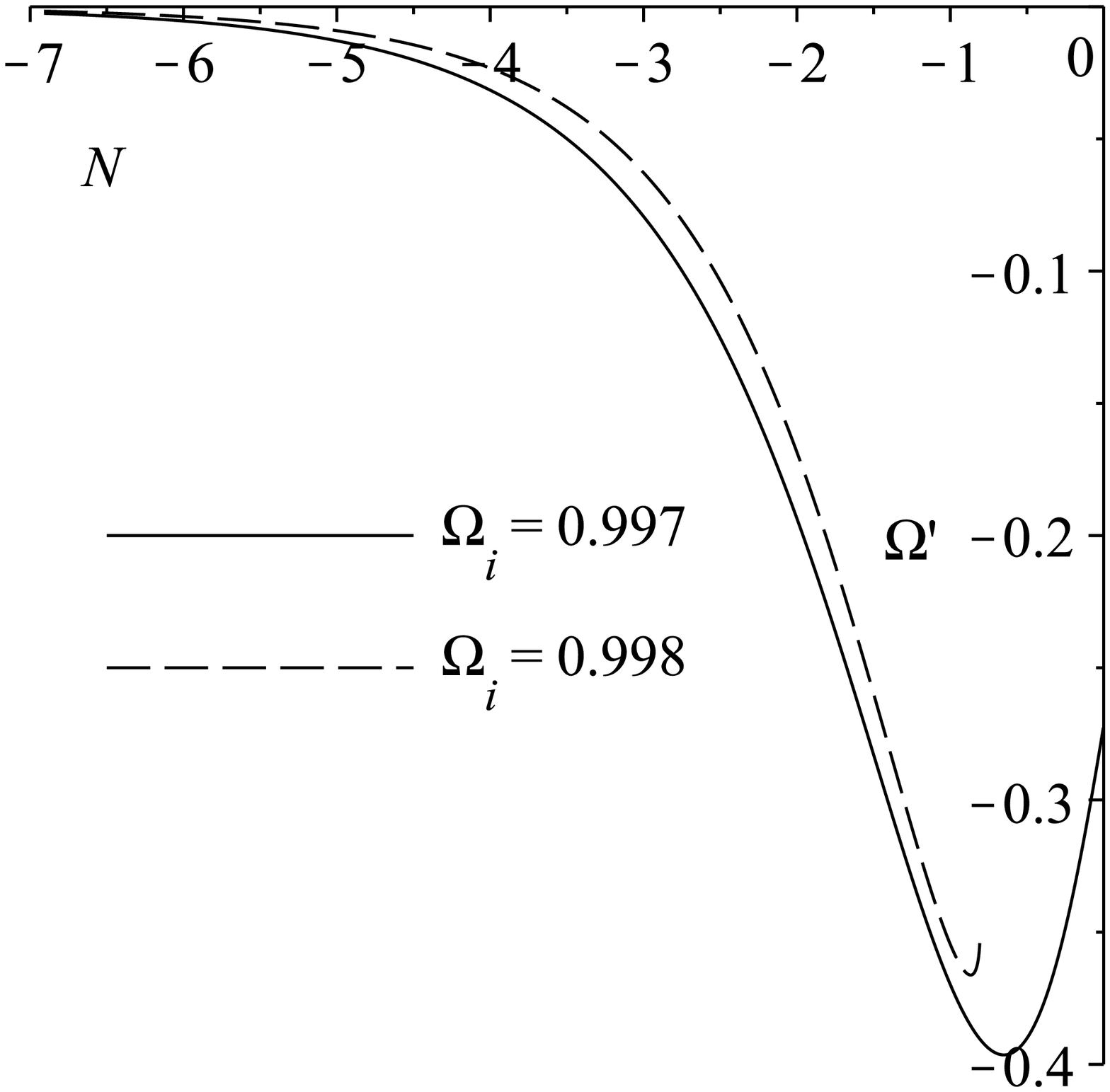}
 \includegraphics[width=\columnwidth]{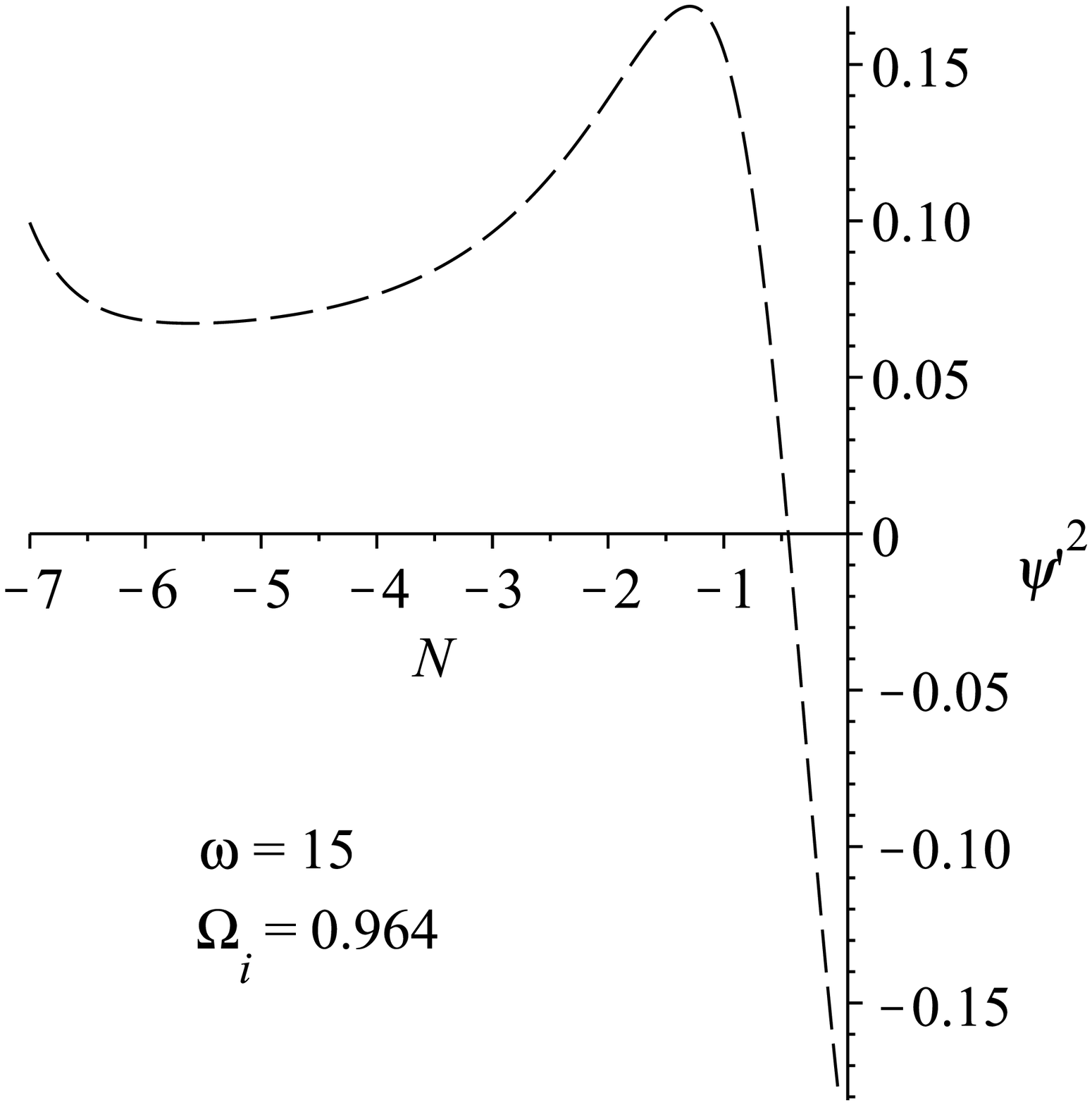}
 \includegraphics[width=\columnwidth]{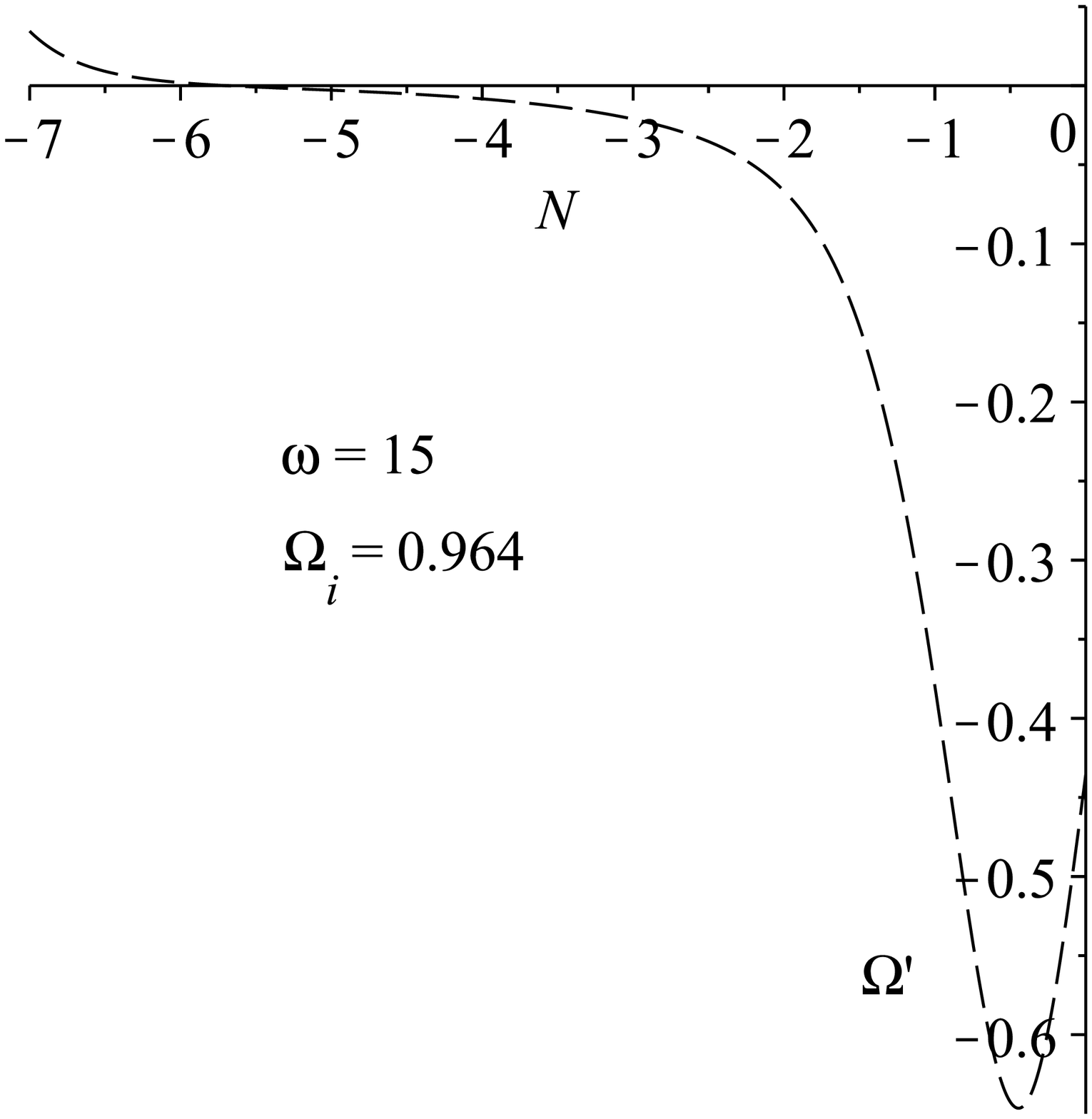}
 \includegraphics[width=\columnwidth]{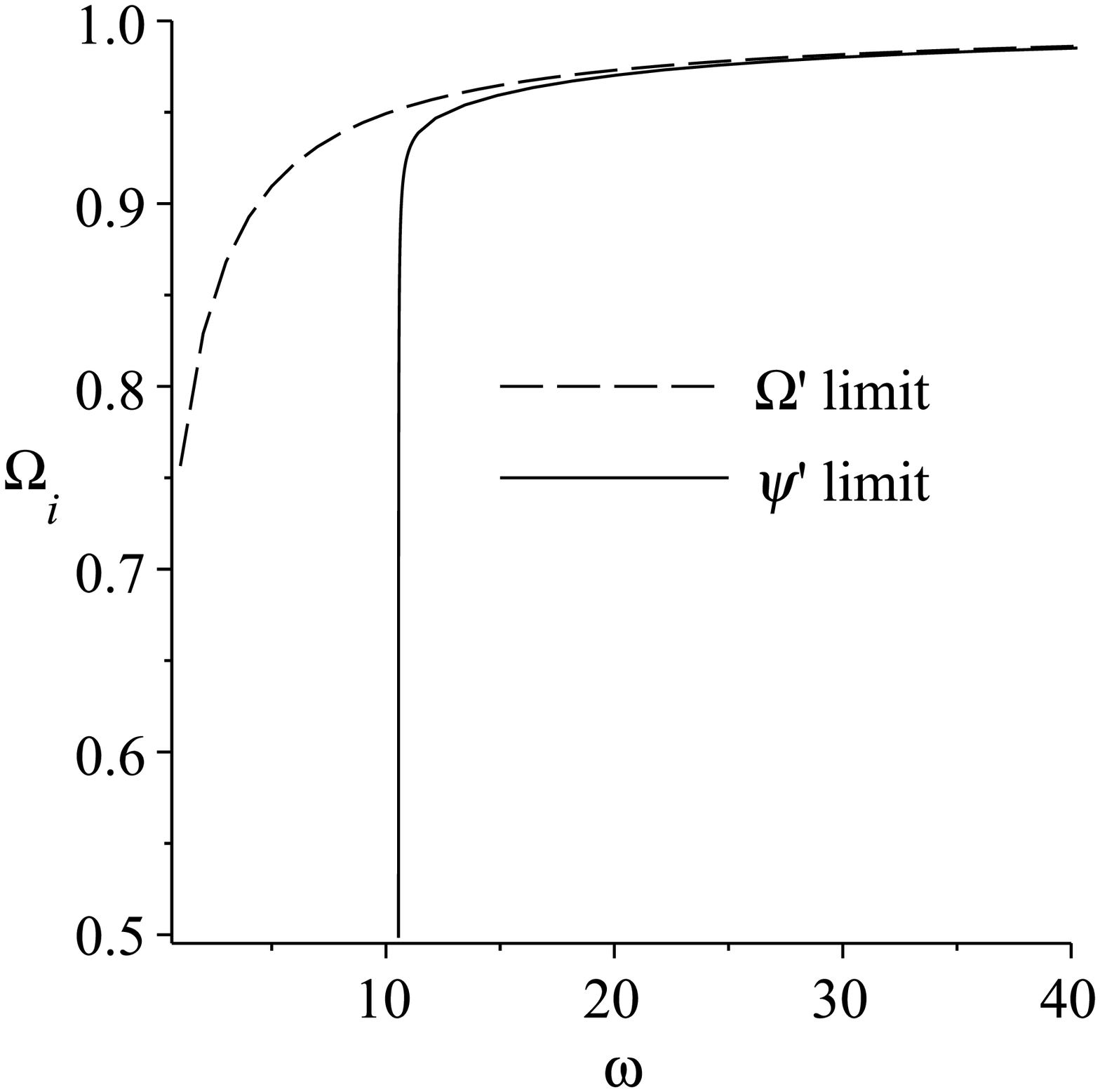}
 \caption{Top left: solutions of $\Omega'$ in the IDE/DGP setup with initial conditions either side of the limit in $\Omega_i$. Top right: an example of $\psi'^2$ becoming negative before the present
day in the IDE/STT setup. Bottom left: solutions of $\Omega'$ in the IDE/STT setup continuing beyond the present day, before which $\psi'^2$ has become negative. Bottom right: $\psi'$ and $\Omega'$
limits on $\Omega_i$ as a function of $\omega$ for the IDE/STT system.\label{iclimits}}
 \end{figure*}
\section{Limit on $\Omega_i$ in the DGP example}\label{dgpiclimit}
For the matched IDE/DGP setup studied in Section~\ref{DGP} there is a limit on how close $\Omega_i$ can be set to $\tilde\Omega_i$. The
differential equation for $\Omega$ from \cite{wei08} which matches the DGP and IDE growth histories is,
\be\kappa Q \phi'\delta'=\frac32\delta\left[\left(1+\frac1{3\beta}\right)\tilde\Omega-(1+2Q^2)\Omega\right],\label{dgpmatch}\ee
where $\kappa^2=8\pi G$, $\phi$ is the IDE scalar field, the function $\beta=-(1+\tilde\Omega^2)/(1-\tilde\Omega^2)$ and $Q$ is the coupling function expressed by,
\be Q^2=\frac{\left(3+2\frac{H'}H-\frac{\Omega'}{\Omega}\right)^2}{-3\Omega-2\frac{H'}H}\label{qsquared},\ee
where $aH=\ch$. Eq.~(\ref{dgpmatch}) is quadratic in $\Omega'$, so to solve it for $\Omega$ we must first solve it for $\Omega'$, but it is not always the case that
real roots exist. Using the initial conditions specified in~\cite{wei08} the solutions are initially complex for $\Omega_i\simeq\tilde\Omega_i$. As $\Omega_i$ is decreased the solutions
extend to later times but there are no solutions which reach the present day
for $\Omega_i\gtrsim0.997$. This can be seen in the top left panel of Fig.~\ref{iclimits} where solutions for values of $\Omega_i$ either side of this limit are plotted.

\section{Limit on $\Omega_i$ in the scalar-tensor theory model and the small $\omega$ limit}\label{stlimitappendix}
For the STT setup of Section~\ref{STT} the solutions of the quadratic Eq.~(\ref{matchedeq}) are not initially complex for $\Omega_i\simeq\tilde\Omega_i$ as they are for Eq.~(\ref{dgpmatch}) of
the DGP setup discussed above. A similar solution limit on how close $\Omega_i$ can be set to $\tilde\Omega_i$ does exist however and depends on $\omega$. In addition there is a physical
limit which is reached before this solution limit and prevents the existence of physical IDE counterparts for those STT cases which deviate most greatly from GR. Similar problems have also
been found in studies of parameterised STT models~\cite{nes06,lee11}.

The denominator in Eq.~(\ref{c2}) can be shown to equal $\psi'^2$ using Eq's.~(\ref{idefluid}-\ref{idefriedmann}). This decreases and reaches zero when the universe begins to accelerate and the
$\frac{\ch'}\ch$ term grows faster than the $\Omega$ term decreases. It can then become negative, which would require $\psi'$ to be complex and so we take this as a physical limit. The top right panel
of Fig.~\ref{iclimits} shows this happening before the present day for a particular choice of $\omega$ and $\Omega_i$, while the bottom left panel shows that at the same time
solutions for $\tilde\Omega'$ still exist.

We plot both the $\psi'$ and $\Omega'$ limits in the bottom right panel of Fig.~\ref{iclimits}, showing that the smaller the value of $\omega$, (and so the
greater the deviation from GR), the farther $\Omega_i$ has to be from $\tilde\Omega_i$. The limit beyond which $\psi'$ becomes complex shows that it is not possible to find a matched IDE/STT system
for $\omega\lesssim10$, contrary to the statement in~\cite{wei08} that for any given MG model it is always possible to construct a matched IDE model.

Note that it is possible to finetune $\phi_i'$ to be very small and negative so that the $\psi'$ limit is avoided, (too much and the STT universe contracts at late-times). Conversely, taking $\phi_i'$
small and positive shifts the limit to much larger $\omega$ making it impossible to find a physical IDE counterpart for cases with any noticable deviation from GR at all.

The reason that the derivative of the IDE scalar field does not become complex for the DGP model can be seen from the `Friedmann' equation of~\cite{wei08} where they define,
\be E\equiv\frac{H}{H_0}=\sqrt{\tilde\Omega_{0}e^{-3N}+\tilde\Omega_{r_c}}+\sqrt{\tilde\Omega_{r_c}}, \ee
with $\tilde\Omega_{r_c}=0.170$. Differentiating this with respect to $N$ and using $\tilde\Omega=\tilde\Omega_{0}e^{-3N}E^{-2}$ leads to,
\be \frac{E'}{E}=\frac{H'}H=-\frac{1.5\tilde\Omega}{1-\frac{\sqrt{\tilde\Omega_{r_c}}}{E}}.\ee
This quantity varies from about $-1.5\tilde\Omega_m$ at early times when $E$ is large, to roughly $-2.5\tilde\Omega$ at late times as $E\to1$. Since $\Omega_m<\tilde\Omega$ at all times we
therefore find a condition which is true at all times in the IDE/DGP setup,
\be (\kappa\phi')^2=-3\Omega_m-2\frac{H'}{H}>0.\ee
\bibliography{mgarticlerefs}
\end{document}